\documentclass{PoS}
\usepackage{url}
\usepackage{amsmath,amssymb}
\usepackage{dsfont}

\newcommand{\gev}{\mathrm{GeV}}

\def\mcL{{\mathcal L}}

\newcommand{\fps}{F_{\rm{PS}}}

\newcommand{\LNP}{\Lambda_{\rm NP}}
\newcommand{\LEW}{\Lambda_{\rm EW}}
\newcommand{\LUV}{\Lambda_{\rm UV}}

\def\fig#1{Fig.~\ref{#1}}

\title{Composite electroweak sectors on the lattice}

\ShortTitle{Composite electroweak sectors on the lattice} 

\author{\speaker{Vincent Drach}\\ 
        School of Engineering, Computing and Mathematics,   University of Plymouth\\
        2-5 Kirkby Place, Drake Circus, PL4 8AA Plymouth, United Kingdom\\
        E-mail: \email{vincent.drach@plymouth.ac.uk}}

\abstract{In the post-Higgs discovery era, the primary goal of the
  Large Hadron collider is to discover new physics Beyond the Standard
  Model. One fundamental question is does new beyond the Standard Model
composite dynamics provides the origin of the Higgs field and
potential.   After reviewing the main motivations to consider composite models based  on a new strongly interacting sector, we 
  summarise the efforts of the lattice community to investigate the
  viability of models featuring a composite Higgs sector. We argue
  that first principle calculations  are necessary in view of the
  fast improvements in accuracy of experimental measurements in the
  Higgs sector. We stress the importance for  lattice calculations to provide
 a testing benchmark for non perturbative mechanisms.  It is
 highlighted that the rich phenomenology of non-abelian gauge theories
 raises a number of questions that can be  explored using lattice calculations. First principle
   results therefore provide crucial insights in the theory  landscape
   that could guide the next generation of  Composite Higgs models.}

\FullConference{37th International Symposium on Lattice Field Theory - Lattice2019\\
		16-22 June 2019\\
		Wuhan, China}

\begin{document}

\section{Introduction}
\begin{figure}[h!t]
\begin{center}
\includegraphics[scale=0.5]{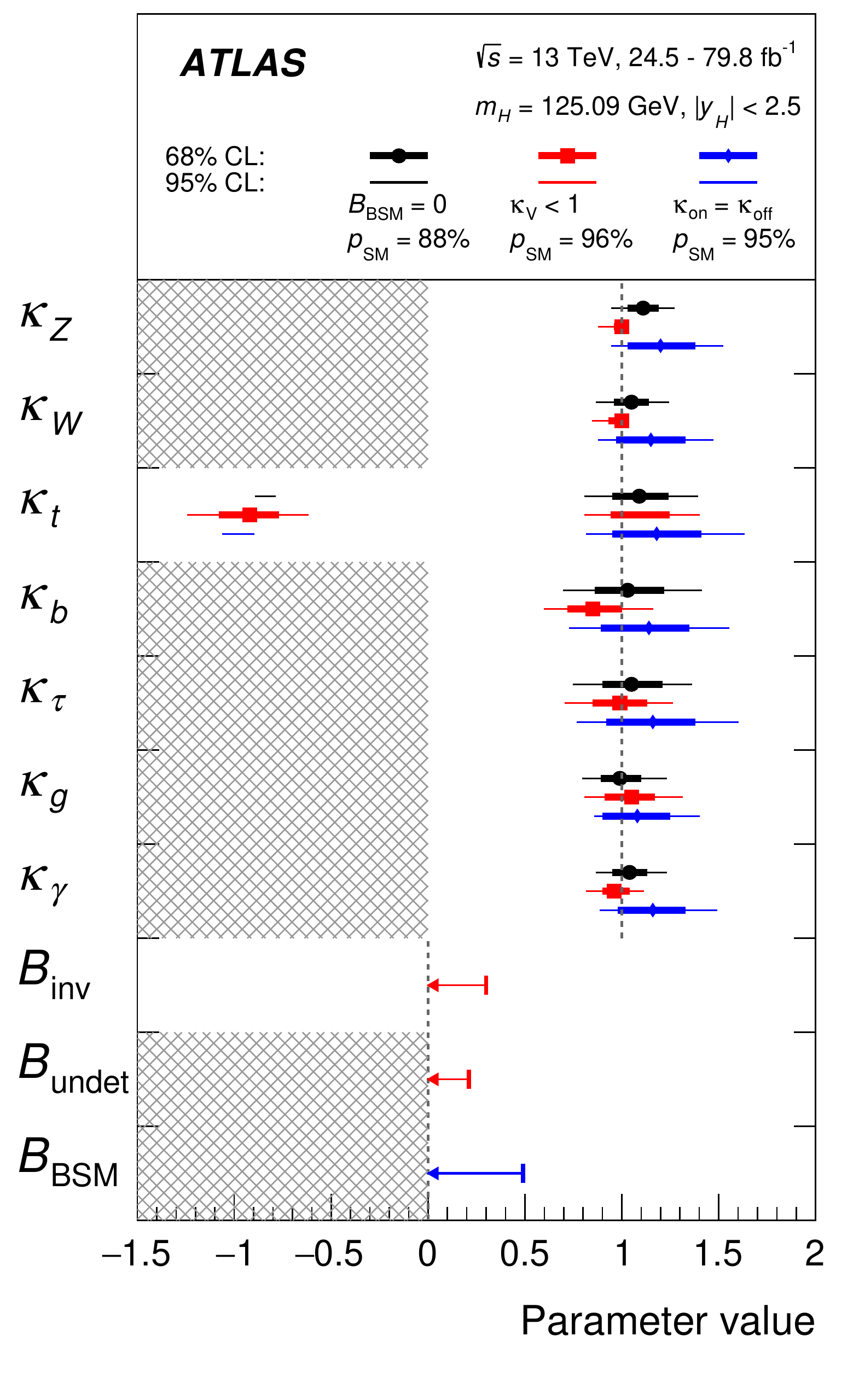}
\caption{\emph{Best-fit values and uncertainties for Higgs boson
    coupling modifiers per particle type with effective photon and
    gluon couplings assuming no BSM contributions (black). More
    details can be found in \cite{Aad:2019mbh}. }}
\label{fig:higgs_coupling}
\end{center}
\end{figure}

The success of the Standard Model (SM) of particle physics to describe
a huge amount of experimental results is undeniable. The
discovery of the Higgs boson in 2012 by the ATLAS and CMS experiments\cite{Aad:2012tfa,Chatrchyan:2012xdj},
after an endeavour that started  half a century ago, marks the
beginning of a new era where finding deviations from the Standard Model
 is crucial to answer fundamental questions about our Universe. A profusion of approaches to the physics
Beyond the Standard Model (BSM) must be explored  and scrutinised in
view of the latest experimental results. Bearing this in mind
 the Higgs discovery is undoubtedly a major step in our
understanding of the interactions at the fundamental level, and
confirms our effective understanding of the origin of mass at the electroweak
scale.  Since 2012,
the combined measurements of the Higgs' mass have been improved to the
subpercent accuracy and read $m_H = 124.97 \pm 0.24 (\rm{Total} \pm
0.16 (\rm{Stat. only}) ~\gev$\cite{Aaboud:2018wps}. The on-going efforts to
determine quantum numbers of the Higgs greatly favoured the ones of the Standard Model,
\textit{i.e} a CP-even scalar particle\cite{Khachatryan:2014kca,Aad:2015mxa}.  A lot of attention has also
been given to the tests investigating the coupling of the
Higgs boson to the fermions and gauge bosons.  Parametrising the
strength of the coupling between the Higgs and vector bosons (V) or
fermions (F), using  the ratio $\kappa_{i=V,F} =
\frac{g_{Hi}}{g^{\rm{SM}}_{Hi}}$, the Higgs coupling to gauge bosons
is measured with a  $10-15\%$ error while the Higgs coupling to the
third generation of fermions is currently measured  with a $20$ to
$30\%$ error\cite{Aad:2019mbh} ($68\%$ CL). These results are
summarised in \fig{fig:higgs_coupling}. Investigations of the Higgs potential remain extremely
challenging at the LHC, but a first bound on the Higgs cubic
self-coupling has been derived by considering
the coupling contribution at EW contributions at
NLO\cite{ATLAS:2019akj}. So far, all the current experimental results suggest a very Standard Model
like Higgs boson.

The experimental results  put severe constraints on new physics
scenarios and in particular, in the context of composite Higgs models
reviewed here, the precision  of the Higgs mass determination
is such that it should not be ignored as a probe to rule out scenario
involving a non-standard Higgs. Until now, the coupling measurements
$\kappa_{V,F}$ provide less stringent constraints.  While  the
experimental results are constantly pushing the limit of the
predictive power of the Standard Model, numerous experimental evidence and
theoretical puzzles that calls for BSM physics are still lacking an explanation.

It is well-known that a number of theoretical and
phenomenological facts question our understanding of the
interactions at the fundamental level and calls for New Physics beyond the Standard Model.
From the theoretical point of view the naturalness problem, the
hierarchy problem or  the strong CP problem  have attracted a lot of attention. From the phenomenological point of view, the Dark
Matter density, the neutrino masses, or  the origin of the
Matter-Antimatter asymmetry, are examples of limitations of the Standard Model. Over the years a large number of mechanisms and theories
have been proposed to explain some or several of these issues and are
under investigations.

Composite Models based on gauge theories with a number of
fermions in various irreducible representations of the gauge group, are
particularly interesting because they are known to generate
scales dynamically and evade the naturalness problem. The fact
that they exhibit a non trivial non perturbative dynamics at low
energy is also source of a rich phenomenology that can be exploited
to build extensions of the Standard Model.  Composite Models face
the challenge that robust quantitative predictions require to resort
to expensive lattice simulations to ultimately compare with
experimental data. Lattice calculations can furthermore provide inputs to
guide model builders, to test mechanisms, and to suggest new
experimental signatures that are sensitive to the underlying dynamics.
Over the years many scenarios relying on a new strong dynamics BSM
composite models have been proposed and received considerable
attention. From technicolor models, to walking technicolor and to Pseudo-Nambu Goldstone Composite
Higgs, the idea of confining new fundamental degrees of freedom into
the known particles is among the most intriguing possibilities to address the flaws of the Standard Model.

Although compositeness can be used in other context, for instance in
the context of Dark Matter,  we limit ourselves to models related to
the dynamical breaking of the electroweak symmetry.

\clearpage

\section{Model building: Pseudo-Nambu Goldstone Composite Higgs
  and Pseudo-Dilaton Higgs}

\begin{figure}[h!t]
\begin{center}
\includegraphics[scale=0.4]{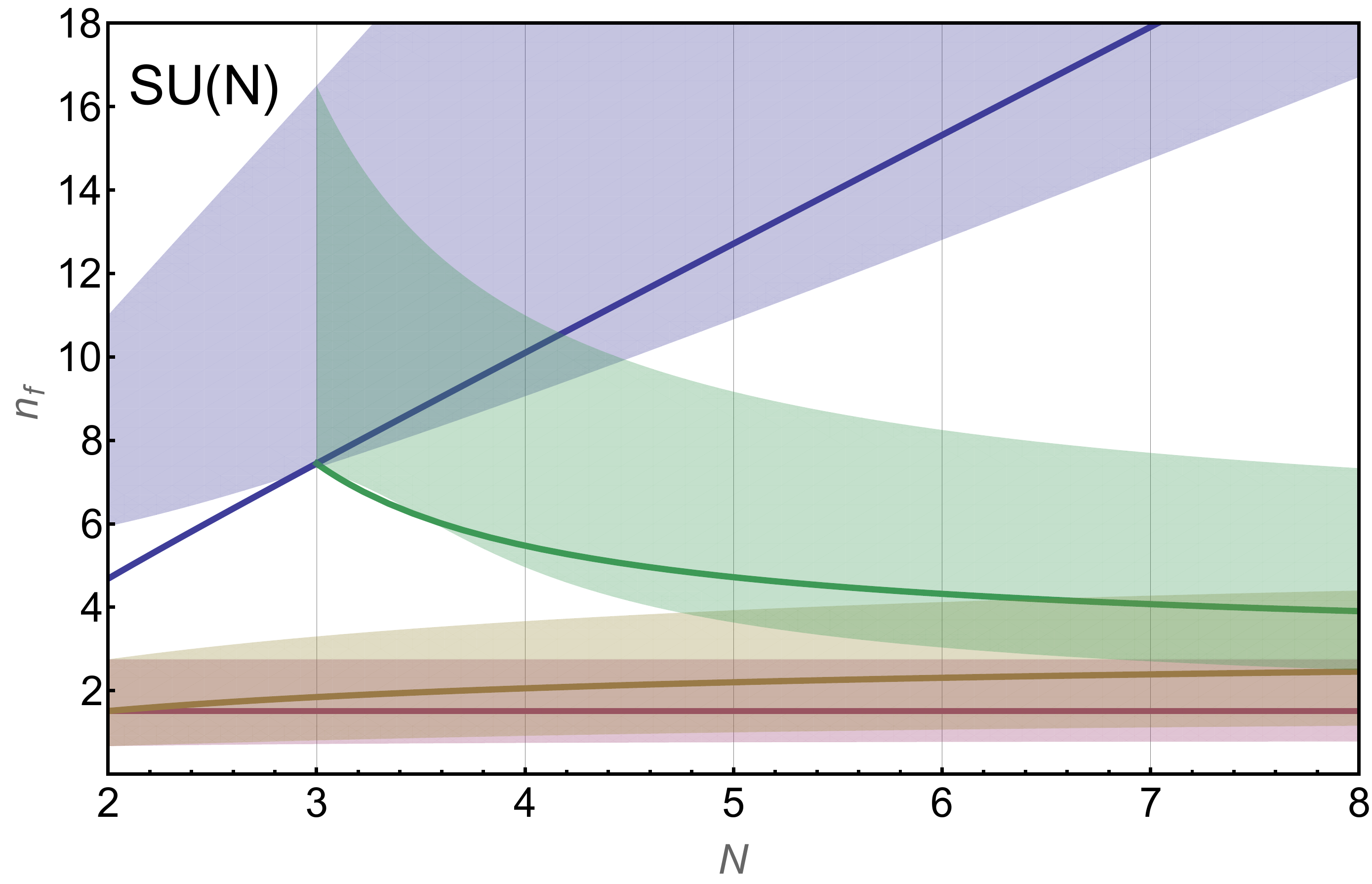}
\caption{\emph{Pertubative (4-loops $\bar{MS}$) calculation of the conformal window for $SU(N)$ groups in the fundamental representation (upper light-blue), two-index antisymmetric (next to the highest light-green), two-index symmetric (third window from the top light-brown) and finally the adjoint representation (bottom light-pink)\cite{Pica:2010xq}. }}
\label{fig:conf_wind_pert}
\end{center}
\end{figure}

Broadly speaking, the underlying theories that are used to design models featuring
a composite Higgs are non abelian gauge theories parametrised
by a a gauge group $G$, and a number  $N_f$ of Dirac massless
fermions in a representation $R$. Depending on the number of fermions and on the gauge group, the large
distance behaviour of the theory is expected to change drastically.
While for $N_f$ large enough, it is known that asymptotic
freedom is lost \cite{Banks:1981nn},  for small enough $N_c$ and $N_f$, for instance in QCD,
asymptotic freedom, confinement and spontaneous chiral symmetry
breaking are expected to occur. In between this two regimes, it is
expected that a conformal window exists, where the $\beta$-function
exhibit an infrared fixed point (IRFP) for the gauge coupling
$g^2$. In that regime, the theory is conformal
at long distances, and all the states of the theory are therefore
massless. We refer the reader to  \fig{fig:conf_wind_pert} for a summary of the
perturbative expectations\cite{Pica:2010xq}. The determination of the
boundaries of the conformal window is however a non-perturbative question,
crucial for model building, that can be studied using lattice field theory..  A brief update on the latest
lattice calculations that address that issue will be discussed at a later stage.  Just below, the critical
number of flavour for which the theory becomes conformal, the system
might develop interesting features for model building, among which the possibility that
the lightest scalar bound state  becomes a pseudo-nambu Goldstone boson associated to
the spontaneous breaking of dilation symmetry, which mass
is controlled by the distance to the conformal window. Lattice
studies are underway to determine if this is occurring or not, but
a number of challenges render the conclusions unclear at this stage.
 
In this section, we focus on the  model building
point of view. We will assume that we have at our disposal a theory
featuring the following properties. First, the theory features  asymptotic
freedom, confinement and spontaneuous chiral  symmetry
breaking. Second, the mass of the  scalar state is close to the spontaneous symmetry breaking
scale $\fps$ in the chiral limit. This last assumption is related to
to the possibility that  the scalar state is a pseudo-nambu Goldstone
boson associated with restoration of the conformal symmetry when $N_f$
is increased. 

In the following, pseudo Nambu Goldstone Bosons (pNGB) will refer to the
degrees of freedom associated with the spontaneous symmetry breaking
of chiral symmetry, while pseudo-Dilaton will refer to the light
scalar state associated to the closeness of the lower bound of the
conformal window.
The pNGB Composite Higgs and pseudo-Dilaton Composite Higgs scenarios share common
features that we start by discussing. The scale of New Physics ($\rm{NP}$) will be referred as
$\LNP$. The Electroweak  ($\rm{EW}$) scale will be denoted $\LEW$.
Models of composite Higgs can be summarised as follow. 
At the New Physics scale, assume that the physics is described by the
effective Lagrangian of the Standard Model without a Higgs sector and
with massless fermions augmented by the Lagrangian of a carefully chosen Strongly
Coupled Theory $\mcL_{\rm SCT}$ which confinement scale is given by
$\LNP$.  At the energy $\LEW \ll \LNP$, the new strongly interacting
sector is replaced by its low energy effective field  theory (EFT), exactly like
QCD is described by chiral perturbative theory at low energy,  so that
the effective Lagrangian is the one of the Standard Model with
massless fermions plus corrections suppressed by powers of $1/\LNP$. 
The origin of the mass of the weak bosons and of the Higgs mass will
differ in the two scenarios and will be discussed below. Fermion masses pose a problem common to all these scenario that we
will discuss later on.  Note that the Lagrangian at the
scale $\LEW$ is dictated by the chiral symmetry
breaking pattern and the quantum numbers of the bound states of the
theory. This explains why many effective models can be studied without even specifying an
underlying strongly interacting sector at $\LNP$. When a gauge theory
$\mcL_{\rm SCT}$ is known in terms of its underlying fermions content,
the model is said to have a UV completion. The choice of the quantum number of the underlying fermions is crucial
to engineer a realisation such a scenario.

We now turn our attention to the differences between the two scenarios
namely PNGB Composite Higgs and the pseudo-dilaton Composite Higgs scenarios.

In the PNGB Composite Higgs case, the underlying gauge theory
$\mcL_{\rm SCT}$ is assumed to have a global symmetry group $G_F$
spontaneously broken down to a subgroup $H_F$.  To preserve the
custodial symmetry $G_{\rm cust} = SU(2)_L \times SU(2)_R$ of the
Standard Model, the quotient group $G_F/H_F$ must be such that $H_F
\supseteq G_{\rm cust}$ and the quantum numbers of the underlying
fermions should be chosen so
that at least one of the Goldstone bosons have the quantum numbers of
the Higgs particle.  The Higgs particle being identified with a
Nambu-Goldstone boson is naturally light \cite{Kaplan:1983fs,Kaplan:1983sm,Banks:1984gj,Georgi:1984ef,Georgi:1984af,Dugan:1984hq}. 

The effective theory at the scale $\LEW$ describes massless SM
fermions. The Higgs mass is generated
by the EW interaction exactly like the pions would acquire
a mass if the electromagnetic interaction is switched on in massless QCD.  A vacuum expectation for the Higgs field is not generated, and
electroweak symmetry remains unbroken. To trigger EW symmetry breaking
the interactions with the SM fermions must be taken into account at
the scale $\LEW$.  Depending on the full Higgs potential,  the Higgs
potential might trigger electroweak symmetry breaking (EWSB). We
assume the potential to be such that EWSB occurs, and we will denote
$V$ the minimum of the potential.  Then, setting the scale so that $\fps \sin{V/\fps}  = v_{\rm EW} = 246
\gev$ where $\fps$ is the pseudoscalar decay constant of the strongly
interacting sector, the mass of the vector bosons  is by construction
the Standard Model one at tree-level.  Furthermore it can be shown that $\kappa_{V,F} = 1 +\mathcal{O}\left(\xi\right)$ where $\xi = \left(v_{\rm EW}
  /\fps\right)^2 $ which therefore guarantees small deviations from the
SM couplings if the so-called vacuum misalignement $\xi$ is small.

In the near-conformal framework, the scale is set by $\fps =  v_{\rm EW} = 246
\gev$ and the EW quantum numbers are chosen so that $m_\sigma = m_H$
and so that  Goldstone bosons become the longitudinal degrees of
freedom of the vector bosons (like in a technicolor scenario). While
in the case of the pNGB scenario, the low energy effective
theory can be systematically written, the question of the existence of
an effective theory describing a light scalar is very challenging and
has stemmed considerable efforts by the community. 

The question of the fermion mass generation, and in a first step, of
the heaviest fermion, the top quark, is a long-standing issue. Several mechanisms
have been proposed. Typically, they require a new sector at an even higher
energy $\LUV$, this new sector typically generates
new effective operators at the scale $\LNP$. Two classes of models have
been proposed: 
\begin{itemize}
\item models that effectively generate operator of the form
$\frac{1}{\LUV^2} q \bar{q} O_{\rm{SCT}}$ , where $q$ stands for the a Standard Model
quark field and $O_{\rm{SCT}}$  a bilinear operator of the new
fermions fields at $\LNP$.  Typically, nothing prevent effective
operators that generate Flavour
Changing Neutral Current which are tightly constrained
by the experiments. Introducing fermion masses into these models is
therefore a challenge.
\item models that effectively generate operator of the form
$\frac{1}{\LUV^{\rm{dim}(O) - 1}}\bar{q}^a O^a_{\rm{SCT}}$, where $a$ stands for a $SU(3)_c$ colour
indices. This scenario is referred to as Partial compositeness
mechanism, introduced in \cite{Kaplan:1991dc}, and requires that the
strongly interacting sector at $\LNP$ to have QCD charged bound
states. Candidate UV completions for partial compositeness scenarios have been
extensively studied, first by
\cite{Ferretti:2013kya,Ferretti:2014qta}.   Other studies have followed \cite{BuarqueFranzosi:2019eee,Gertov:2019yqo}
\end{itemize}

Fermion mass generation in composite scenario is a challenge for
theorists and lattice simulations start be used to test fermion mass
generation mechanisms and  to guide the model building community.

\begin{figure}[h!t]
\begin{center}
\includegraphics[scale=0.3]{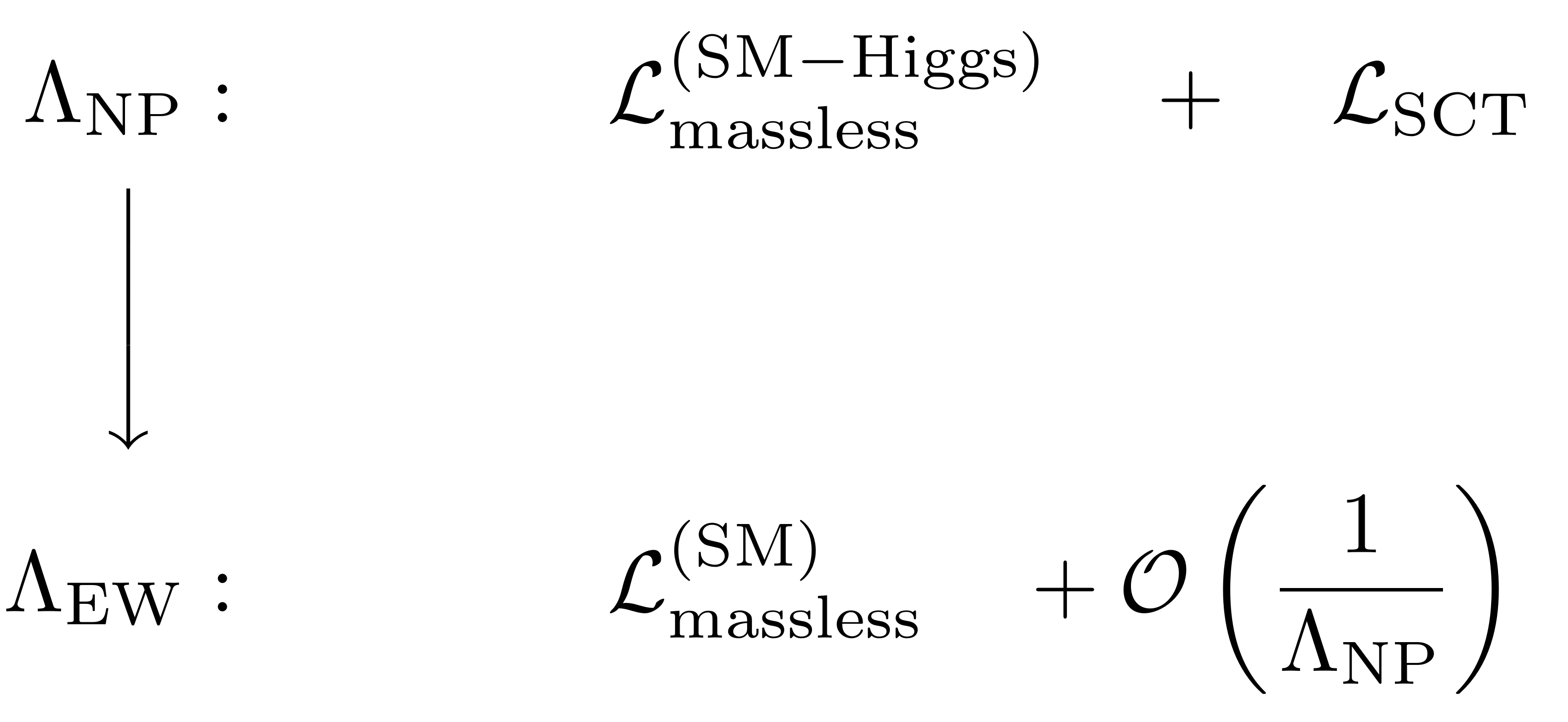}
\caption{\emph{A sketch of  composite  Higgs scenarios with massless
    fermions (at LO).}}
\label{fig:sketch_compositness}
\end{center}
\end{figure}

\newpage
\section{Lattice investigations of near-conformal scenarios}

Many lattice groups are accumulating evidence that near-conformal dynamics
gives rise to a light scalar state.  The focus has been on
models based on $SU(2)$ and $SU(3)$ gauge groups. The lattice
calculations of the lightest scalar singlet are demanding because of
the costly disconnected contributions associated with the scalar
bilinear interpolating field. 

The interpretation of lattice calculation is difficult
because the prediction of the spectrum in the chiral limit depends strongly of the
choice of the effective theory describing the low-lying states. For instance, chiral
perturbation theory based on the chiral lagrangian is expected to receive large contributions from the
scalar state. Models and effective theories
have been designed and consistency checks are now being performed to
conclude about their ability to describe the lattice data.

At finite fermion mass, several groups observe signs of chiral
symmetry breaking and of a scalar
states that have a mass
very similar or even below the mass of the Goldstone bosons. This justifies, a
posteriori, the use of two-point functions to obtain the mass of the
state without having to consider a full-fledged
finite volume calculation of the two Goldstone bosons scattering
process. In order to improve our understanding of near-conformal gauge
theories, further investigations both at the numerical level
by reducing the systematics and at the theoretical level
by progressing in our understanding of the effective description are required.

\begin{figure}[h!]
\begin{center}
\includegraphics[scale=0.6]{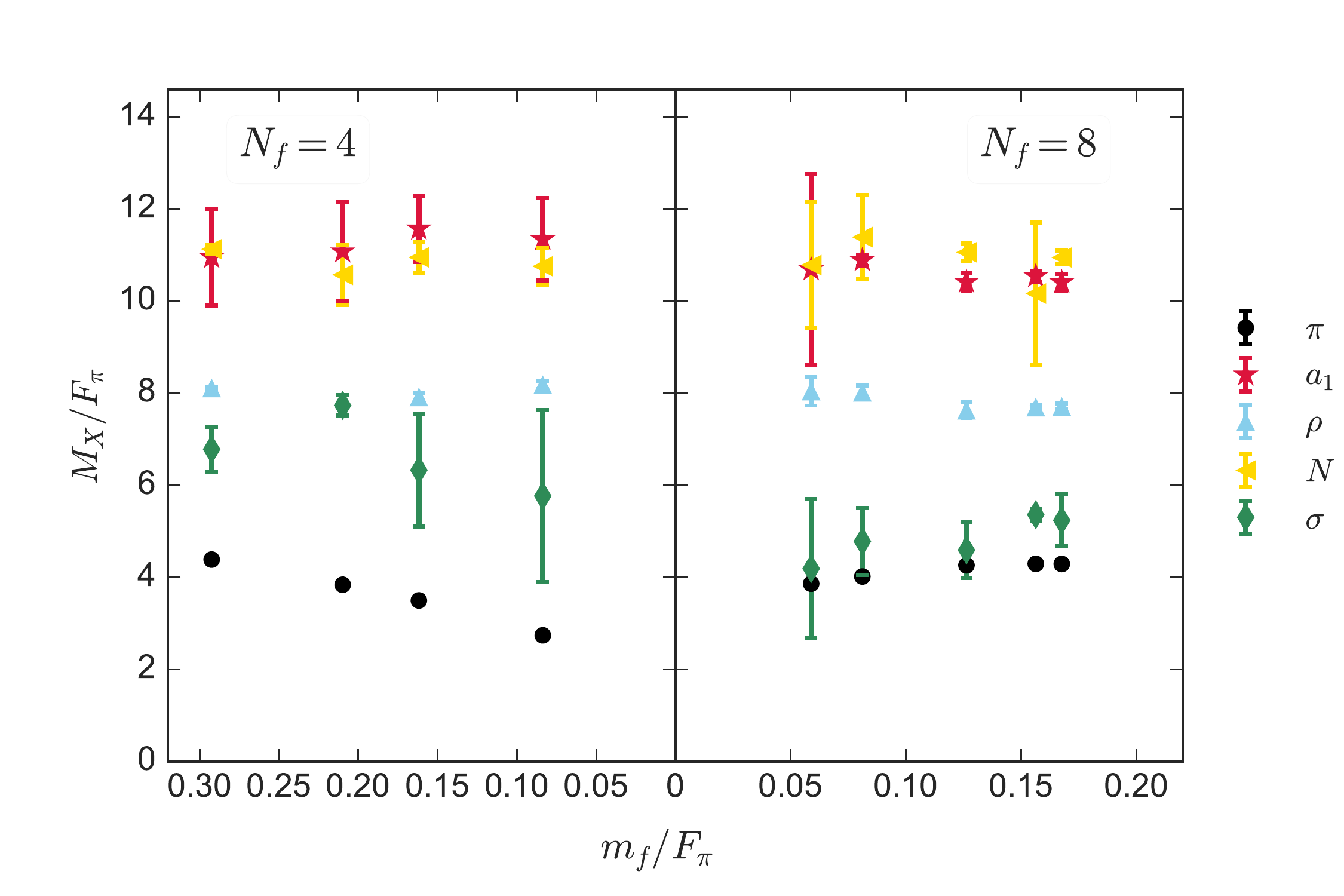}
\caption{\emph{Comparison of the spectroscopy  of $SU(3)$ gauge theory
    for $N_f = 4$(left) and $N_f = 8$ (right) fundamental fermions with. Hadron masses (vertical axis) and
    the fundamental fermion mass (horizontal axis) are both shown in
    units of the pion decay constant $F_\pi$; the chiral limit $m_f =
    0 $ is    at the center of the plot for both theories. The major
    qualitative difference between the two values of $N_f$ is the
    degeneracy of the light scalar $\sigma$ with the pions at $N_f =
    8$. \cite{Appelquist:2018yqe}}}
\label{fig:LSD}
\end{center}
\end{figure}

To date, the theories that
observe candidates for light scalar states are: $SU(3) $ with $N_f= 8$
fundamental fermions, as illustrated in \fig{fig:LSD}, $SU(3)$ with $N_f=2$ antisymmetric (or sextet)
fermions and also theories are likely to be conformal $SU(2)$ with
$N_f=1$ or $2$ adjoint fermions. 

\subsection{Models and Effective theories of near-conformal dynamics}

 A number of models and effective theories have been
proposed to describe theories close to the lower bound of the
conformal window which include the scalar field as a degree of freedom\cite{Holdom:2017wpj,
  Goldberger:2008zz,Matsuzaki:2013eva,Golterman:2016lsd,Appelquist:2017wcg,Appelquist:2018tyt,
  Golterman:2018mfm,Golterman:2018bpc,DeFloor:2018xrp, Fodor:2020niv}.  The goal of this section is not to review them but
to provide an up-to-date list of the low energy descriptions, and to
highlight their main features.   The discussion of the latest tests of these low energy description
using lattice data is postponed to the next section.

\paragraph{Bound state model:}
{\textit Holdom and Koniuk} consider a Hamlitonian-based bound state model where
the pseudo-scalar, scalar, vector and axial-vectors are included\cite{Holdom:2017wpj}. They
study the relation between the spectrum  and the relevant form
factors.

\paragraph{Chiral perturbation theory with a flavor-singlet scalar:}
the central idea is to augment the chiral Lagrangian by an
iso-singlet scalar in the most general
way\cite{Soto:2011ap,Hansen:2018gck}. The corresponding low energy
description does not rely on the closeness from  the lower bound
of the conformal window, but it thought to be able to capture a
variety of underlying dynamics where the iso-singlet scalar plays a role.

\paragraph{Generalised Linear sigma model}
 based on an approximate 
  infrared conformal invariance, the scalar potential breaks chiral symmetry spontaneously\cite{Appelquist:2018tyt}. 

\paragraph{Effective theory of a pseudo-Nambu-Goldstone boson
  associated to the spontaneous breaking of the conformal
  invariance:}
several authors build on the idea that the scalar state is associated
to the spontaneous breaking of the dilation symmetry. There is a long
history of effective description of the dilaton Lagrangian depending
on the focus of the authors \cite{Goldberger:2008zz,Matsuzaki:2013eva,Golterman:2016cdd,Golterman:2016lsd}.
\textit{Appelquist, Ingoldby and Piai} add Nambu Goldstone bosons associated
to the spontaneous breaking of chiral symmetry to build their effective
lagrangian\cite{Appelquist:2017wcg}. \textit{Golterman and Shamir} derive an effective
theory by introducing a systematic expansion in terms of a parameter controlling the distance to the
lower bound of the conformal window in their Dilation-pion low-energy
effective theory\cite{Golterman:2016lsd}.  The authors also derived a
systematic expansion in which the fermion mass is not small relative
to the confinement scale. In the large-mass regime of the Dilaton low-energy
theory, they have shown that at leading order hyper-scaling
relations are expected \cite{Golterman:2018mfm}.

\paragraph{Complex conformal field theory: }
inspired by the work of \textit{Gorbenko et al.}\cite{Gorbenko:2018ncu,Gorbenko:2018dtm}, and in
particular of their analysis of the two-dimensional $Q-$state Potts model with $Q>4$,
{\textit Kuti} argues that the slow running of the coupling constant
in near conformal theory might be related to the presence of two fixed
points at complex coupling, referred as complex
CFTs\cite{Fodor:2020niv}. He suggests an other avenue to explore other
types of low energy descriptions.

\subsection{Testing low-energy descriptions using lattice results}

Several theories are being extrapolated using the low energy
descriptions discussed in the previous section. The latest findings will be reviewed here.
 
The Lattice Strong Dynamics Collaboration presented new preliminary results including pseudo-scalar
decay constant and mass, the mass of the scalar state and scattering
length of the Goldstone bosons for the $SU(3) $ gauge theory with
$N_f= 8$ fundamental fermions\cite{Fleming}. The results suggest
that the linear Sigma Model EFT describes the data well.  Expressions
for scattering lengths of GBs and the scalar decay constant have been
derived and additional lattice calculations are underway.

The LSD data have also been confronted to a large mass regime of the
Dilaton-EFT \cite{Golterman:2018mfm}. The LSD data
are well described by the hyper-scaling  scaling predicted under the
assumption that the fermion mass is not small compared to the
confinment scale \cite{Golterman:2018bpc}.  The authors
have now shown that the EFT describes the taste-splitting pattern in
the discretisation of the theory with staggered fermions. The observed pattern is indeed very
different from the one observed in staggered lattice QCD \cite{Golterman:2019htd}.  
This fact brings additional evidence  that  the assumptions that
underpin the Dilaton EFT, and in particular its large-mass regime, are
predictive.

\begin{figure}[h!]
\begin{center}
\begin{tabular}{cc}
 \includegraphics[height=6cm]{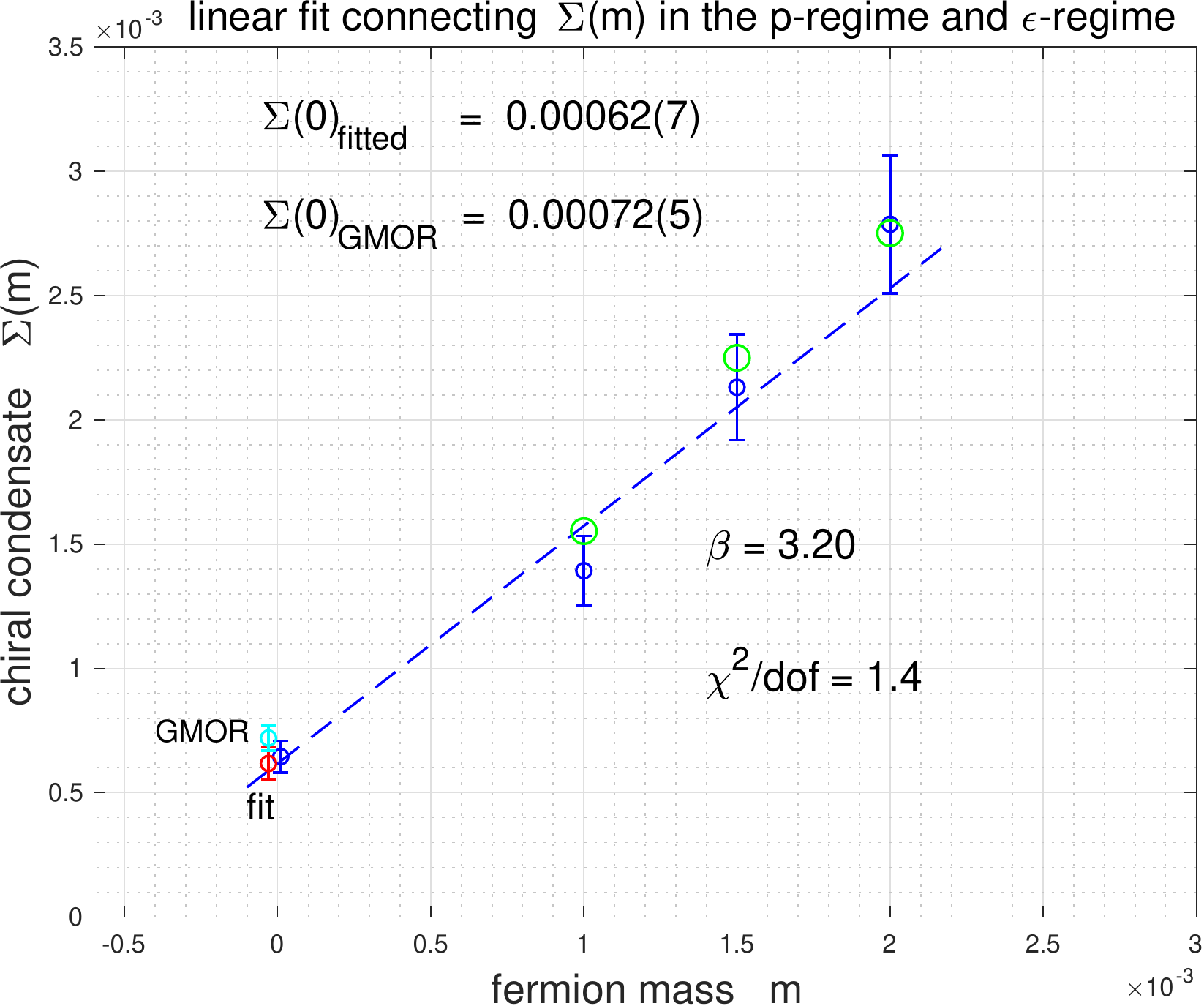}&
 \includegraphics[height=6cm]{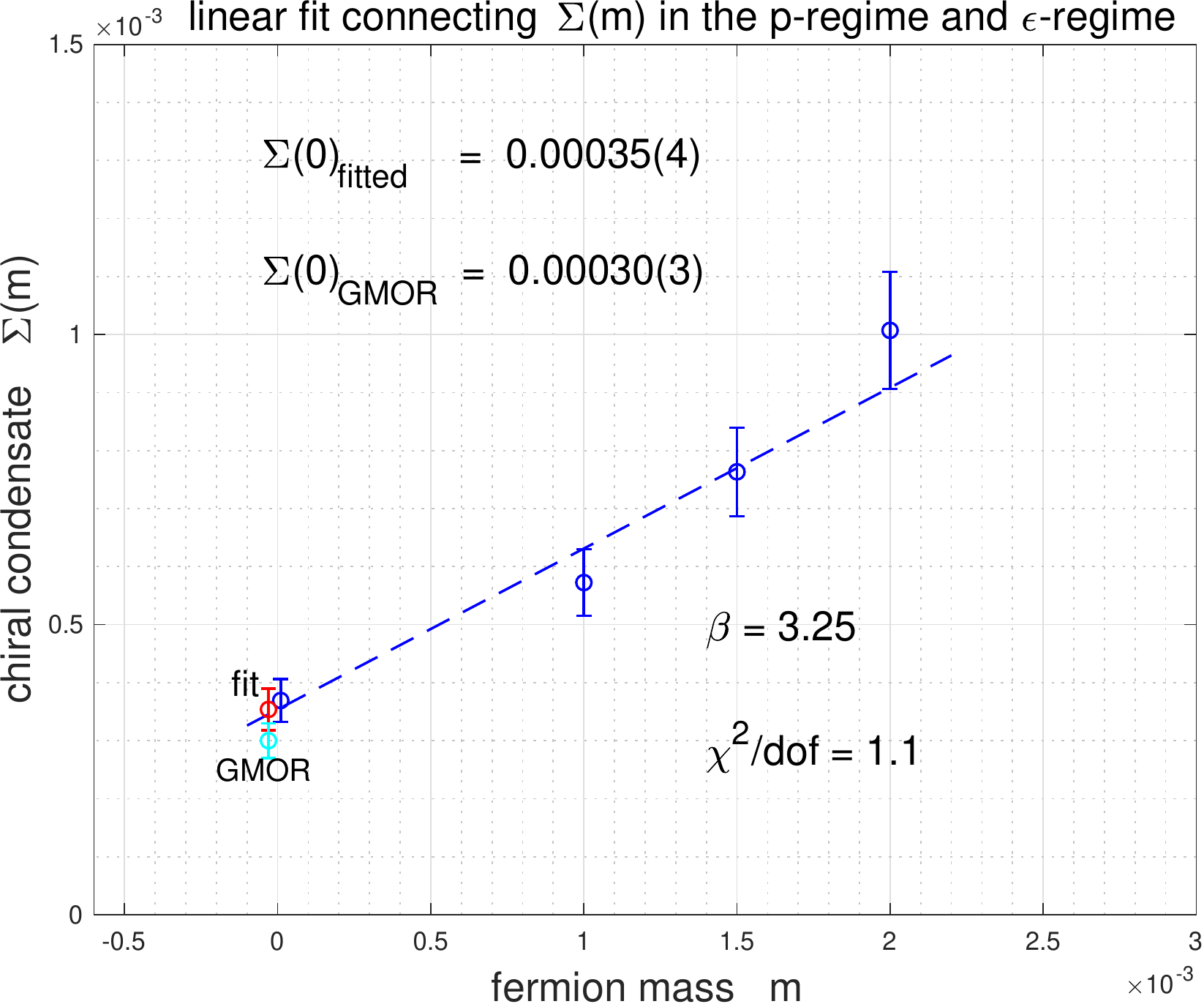}
 \end{tabular}
\caption{\emph{Comparison of the chiral condensate $\Sigma(m)$
  obtained from simulations in the $\epsilon$-regime and by the GMOR
  relation in the $p$-regime for two sextet (Dirac) fermions of
  $SU(3)$ gauge theory. More details can be found in \cite{Fodor:2020niv}.}}
\label{fig:sextet}
\end{center}
\end{figure}

The latest result regarding simulations of two flavours in the symmetric (sextet)
representation of the SU(3) gauge group,  the LatHC collaboration, presented results at two
lattice spacings\cite{Fodor:2020niv} and tested the dilaton hypothesis with two
typical dilaton potentials. Infinite volume extrapolation of the
pseudo-scalar mass and decay constant have been performed and the
authors concluded that the dilaton decay constant over pseudo-scalar
decay constant and the ratio of the dilaton mass over the GBs mass are
very sensitive to the mass of the scalar state. More recently the authors
also used Random Matrix Theory in the $\epsilon$-regime  to
test further the consistency of their dilatonic  Lagrangian. In
\fig{fig:sextet}, the chiral
condensate obtained from the GMOR relation and from Random Matrix Theory in
the $\epsilon$-regime are  compared and show good agreement. 

\subsection{Mass-split models: $SU(3)$ with $4$ light and $6$ heavy fundamental flavours}

Motivated by the experimental fact that the Higgs boson is light and
that no other heavier resonances have been observed, describing a
composite Higgs boson requires a system with a large separation of
scales.  An alternative approach to near-conformal theory is based on mass-split
models which feature a number of light and heavy flavours. Those theories are chirally broken in the IR but conformal in the
ultraviolet (UV). Such a class of model features interesting properties
from the model building point of view but fails to
explain the origin of the two well separated mass scales in a first place.

Mass-split systems can be  used to design models involving a light
dilaton or  Composite Higgs models  based solely on the spontaneous
breaking of chiral symmetry scenarios \cite{Luty:2004ye,Dietrich:2006cm,Vecchi:2015fma,Ferretti:2013kya}.
In the UV, the number of flavours is chosen to lie in the Conformal
Window and the theory is therefore driven by the conformal fixed point. In
the IR, the heavy flavours decouple and the system therefore exhibits
spontaneous chiral symmetry breaking. 

It has been shown that in such mass-split system, dimensionless ratios
exhibit two important features\cite{Brower:2014dfa,Brower:2015owo,Hasenfratz:2016gut}.   First,
dimensionless ratios of physical observables exhibit hyper-scaling
: they are function of the ratio $m_;/m_h$ . Such ratios are therefore
independent of the  mass of the heavy flavours. Second, the authors
argue that the range of slowly evolving (``walking'') coupling
increases when the mass of the heavy flavours is reduced. The walking
region can therefore be tuned at the price of introducing
explicitly two scales.

Mass-split systems have been investigated previously with $4$ light
and $8$ heavy fundamental flavours of $SU(3)$ and it seems that the
lightest scalar is the lightest massive particle in the chiral limit\cite{Brower:2015owo}.

Results for $4$ light and $6$ heavy flavours of $SU(3)$ have been
presented\cite{Witzel:2019oej}. They show that the connected meson
spectrum exhibits hyperscaling. The calculations of the scalar meson
have not been performed yet.

\begin{figure}[tb]
  \centering
 \includegraphics[height=0.23\textheight]{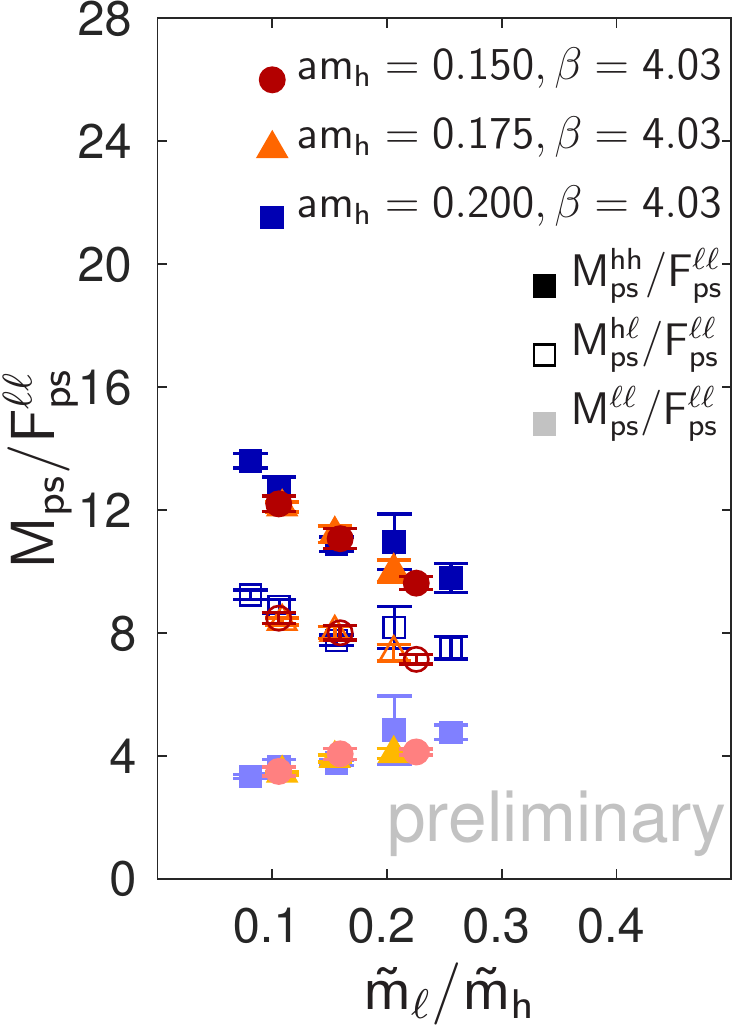}
  \includegraphics[height=0.23\textheight]{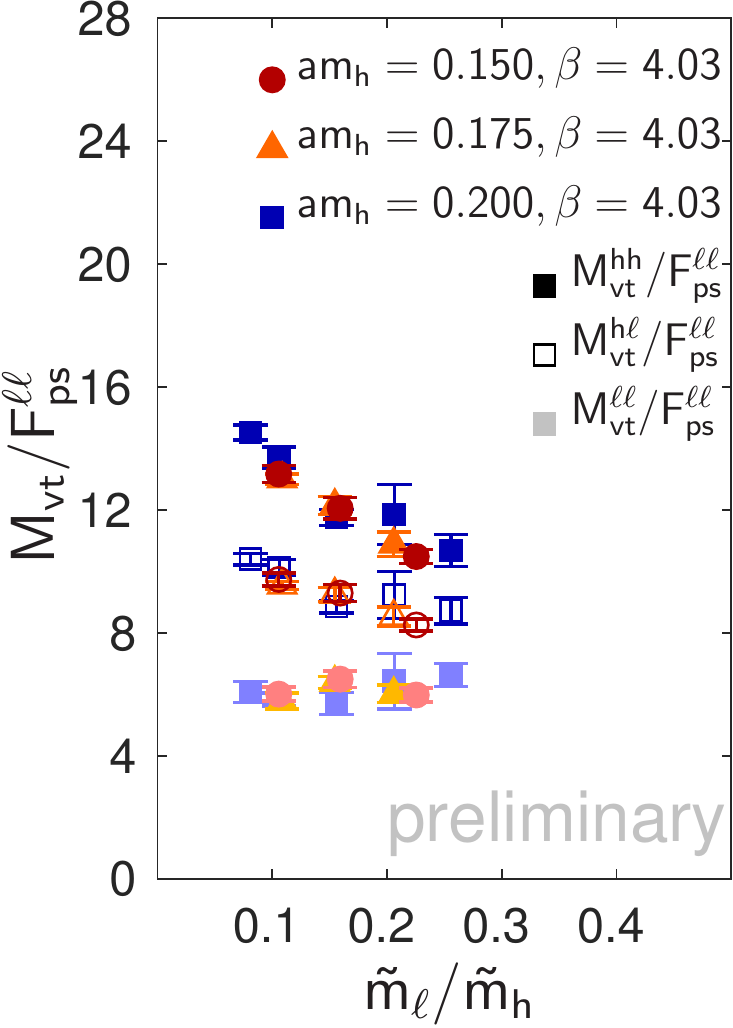}
  \includegraphics[height=0.23\textheight]{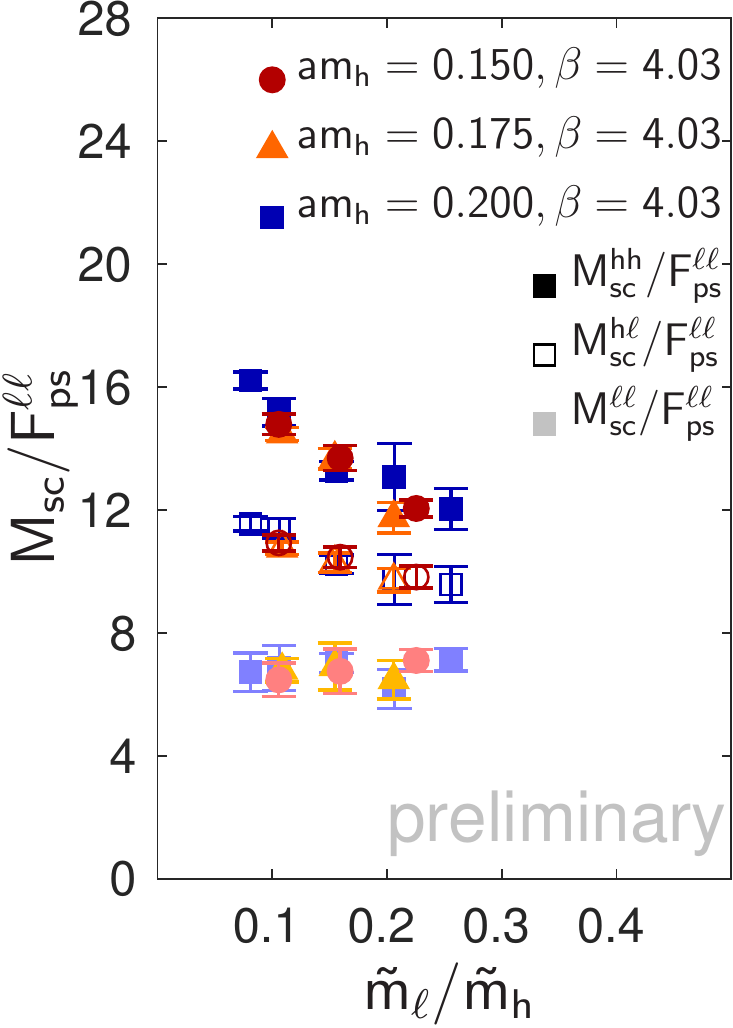}
  \includegraphics[height=0.23\textheight]{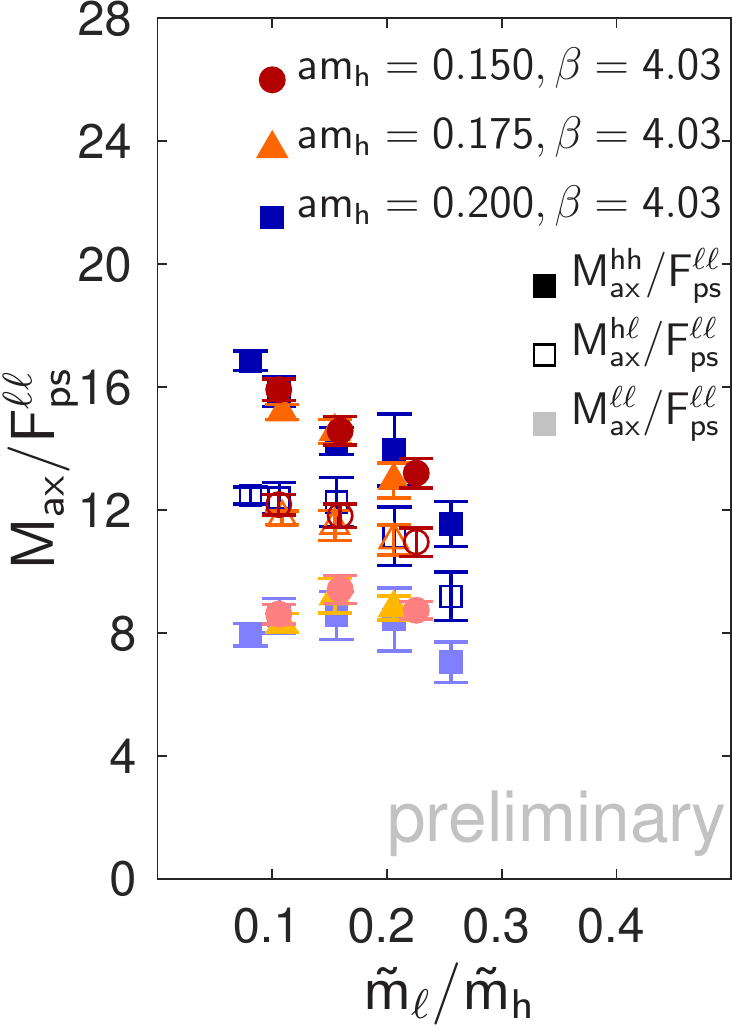}
  \caption{\emph{Low-lying connected meson spectrum obtained from light-light, heavy-light, or heavy-heavy two-
    point correlator functions as a function of the ratio of light
    over heavy flavour mass to highlight hyperscaling\cite{Witzel:2019oej}.}} 
 \label{fig:mass_split}
\end{figure}

\newpage
\section{Lattice investigations of Pseudo-Nambu Goldstone Boson Higgs scenarios}

We review here the questions addressed by lattice collaborations
to test PNGBs scenarios in view of the latest experimental results and to provide first principle
predictions to guide phenomenologists.

The first information that lattice collaborations can provide regarding a UV
completion is its spectrum in isolation of the Standard Model.  A
related question is to determine if the spectrum is similar to the one observed
in QCD or on the contrary rather different. Identifying common
features of different UV completion allow to understand better what
could be the experimental signatures of such models.

Other low energy properties of UV completion in isolation can also
provide insights: form factors, properties of the candidate Higgs
boson, the Higgs potential itself, and scattering properties of the PNGBs are as many possible
interesting observables that can bring essential information in the
search for experimental signatures of a composite electroweak sector at the LHC.

A third class of problems that is addressed is to use lattice calculations as a laboratory to test
mechanisms for the generation of the top quark mass or to pave the way
for a calculation of the Higgs potential.

\subsection{$SU(2) = Sp(2)$ gauge theory with $N_f  =2$ fundamental fermions}

\begin{figure}[h!t]
\begin{center}
\includegraphics[scale=0.5]{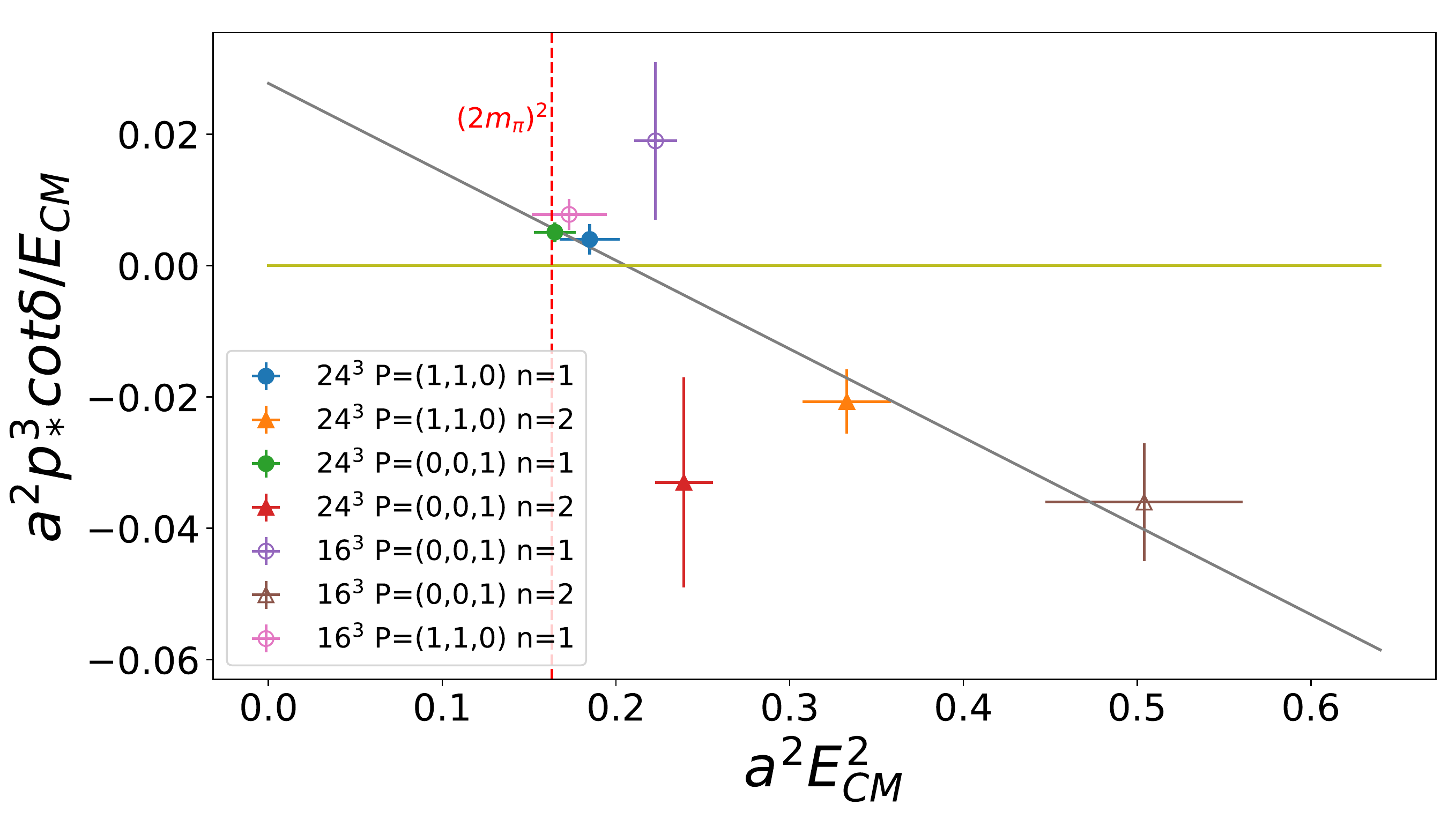}
\caption{\emph{Plot of $a^2 p^{\ast 3} \rm{cot} \delta / E_{\rm
      CM}$ as a function of the
    squared centre-of-mass energy. Points with error bars correspond
    to energy levels from different ensembles and/or total momenta P.
    A linear fit allows to determine the resonance parameters.}}
\label{fig:su2}
\end{center}
\end{figure}

The most minimalistic known UV completion of PNGBs Higgs boson
scenario is based on $SU(2)$ gauge theory with two fermions in the
fundamental representation. The theory is far from the conformal
window and is therefore QCD-like.  Because the fundamental
representation  of $SU(2)$
is pseudo-real, the flavour symmetry of the classical massless theory
is upgraded to $SU(4)$. Classically, the mass term breaks the flavour
symmetry down to $Sp(4)$.  A number of publications have confirmed non-perturbatively that the chiral breaking pattern is $SU(4)
\longrightarrow Sp(4)$
leading to five Goldstone Bosons.  The EW embedding has been proposed
in \cite{Cacciapaglia:2014uja}, and the model is phenomenologically viable \cite{Arbey:2015exa}.
The fact that this theory is used as UV completion by model builders
makes it ideal to explore the underlying dynamics in details.
For the first time, a full non-perturbative calculation of the vector
meson width has been presented\cite{Janowski:2019svg}. The phase-shift
is calculated using  the energies of two-pions in  finite volume and
are related to infinite-volume scattering amplitudes
$S(E)=e^{2i\delta(E)}$  via rigorous L\"uscher's
formalism~\cite{Luscher:1990ux, Rummukainen:1995vs}.  The resonance
parameters $( g_{\rho\pi\pi} , M_\rho$) are obtained from the phase shift $\delta(p)$ using the parametrisation:
  \begin{equation}
  \label{linear_fit}
      \frac{p_*^3 \cot \delta}{E_{CM}} = \frac{6
        \pi}{g_{\rho\pi\pi}^2} \left( M_\rho^2 - E_{CM}^2 \right)\, 
  \end{equation}
as illustrated in \fig{fig:su2}.

 The calculation shows
that the preliminary value of the coupling constant that control the vector meson
width is $g_{\rho\pi\pi} \sim 11(2)$, a value  somewhat larger than in QCD where
$g_{\rho\pi\pi} \sim 6$.   This result does not include chiral or
continuum extrapolations needed to extract the physical value of
$g_{\rho\pi\pi}$.   The number can be compared to the
phenomenological relation obtained using vector meson dominance by
Kawarabayashi \& Suzuki \cite{Kawarabayashi:1966kd} and by Riazuddin
\& Fayyazuddin\cite{Riazuddin:1966sw} (KSRF)  stating that $g^{\rm{KSRF}}_{\rm VPP} =
m_V/\sqrt{2} \fps$. Using numberes from\cite{Arthur:2016dir},  the KSRF estimate of the
coupling read   $g^{\rm{KSRF}}_{\rm VPP}  = 9(2)$. The preliminary
result, obtained at
finite quark mass and finite lattice spacing, suggests that
the KSRF turns out to be satisfied for $SU(2)$ gauge theory with
$N_f=2$ fundamental fermions.
The consequences of such a finding in terms of constraints provided by
the LHC remain to be investigated.

\subsection{$Sp(4)$ gauge theory with $N_f  =2$ fundamental fermions}

The latest results on the on-going research programme started in
\cite{Bennett:2017kga}  focusing on $Sp(4) $with  $N_f  =2$
fundamental fermions have also been presented.  The UV completion share the same chiral symmetry breaking pattern
$SU(4) \rightarrow Sp(4)$ as $SU(2)=Sp(2)$ with two fundamental
flavours and therefore shed light on the dependence of the physical
observables on the gauge dynamics.
 One very interesting aspects of the $Sp(4)$ gauge
 dynamics is that when the two fundamental flavours are supplemented
 by  three antisymmetric fermions, the theory features top-partner
 candidate and can be shown to be a UV completion of  partial
 composite models. More details concerning this line of research can
 be found in a recent paper\cite{Bennett:2019cxd}. Finally, the model is also relevant in the context of
 strongly interacting massive particles (SIMP) as dark matter  candidates\cite{Hochberg:2014kqa}.

\begin{figure}[tb]
  \centering
 \includegraphics[height=0.21\textheight]{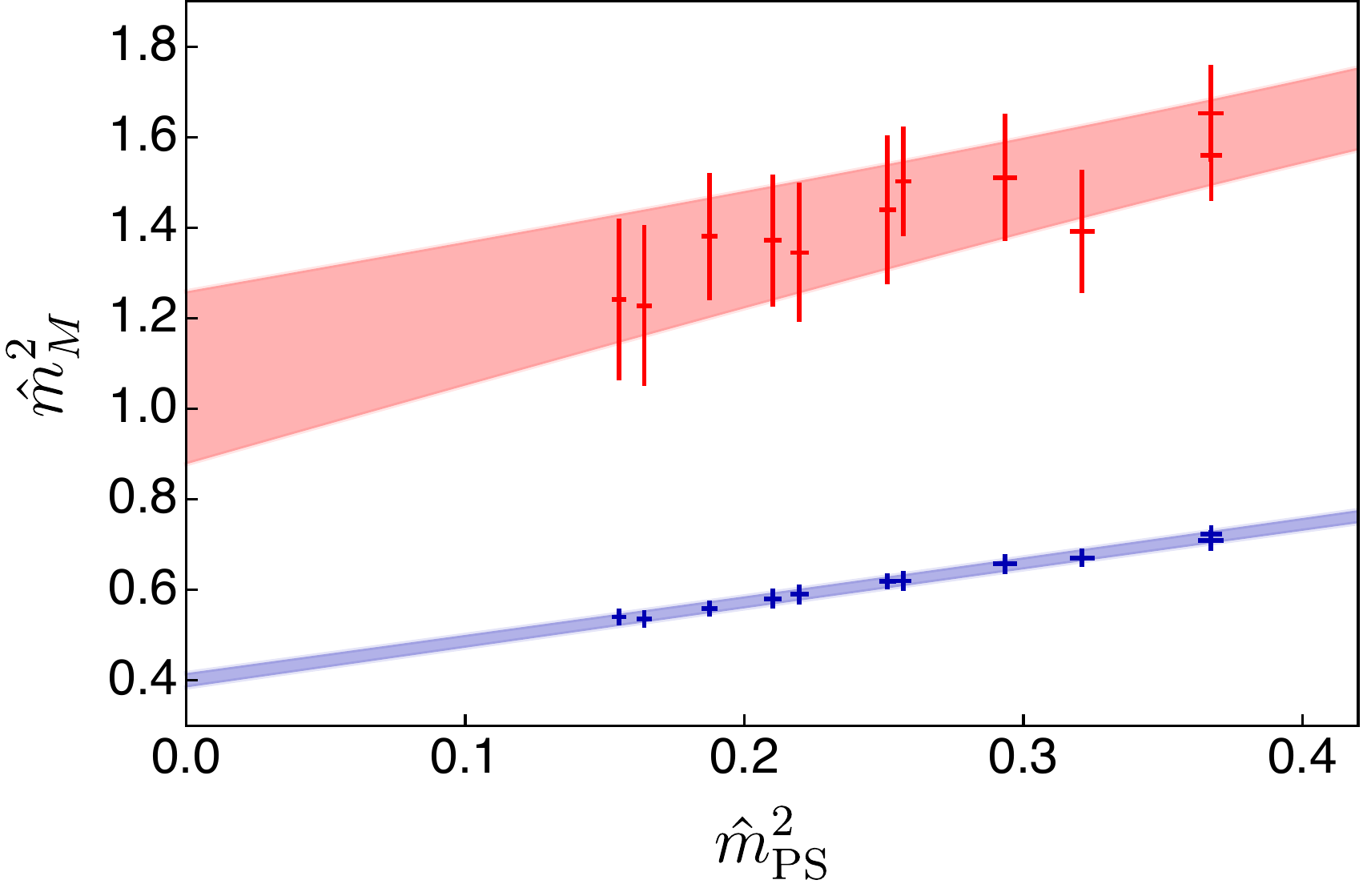}
  \includegraphics[height=0.21\textheight]{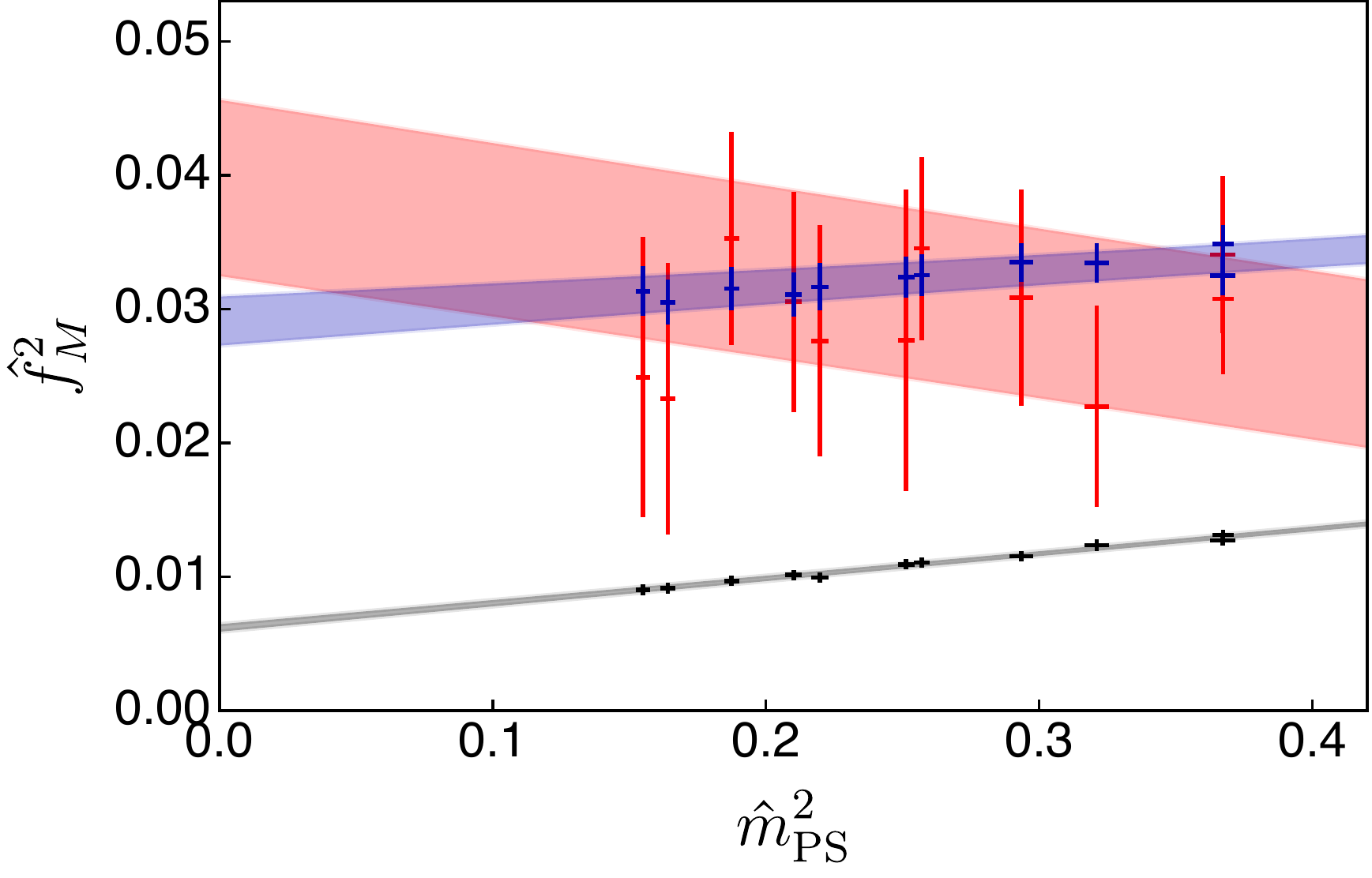}
   \caption{\emph{Continuum-extrapolated meson masses and decay constants squared as a function of the pseudo-scalar mass squared. Black, blue and red colours are for PS, V and AV mesons. The global fit results are denoted by shaded bands with their widths representing the statistical errors.}}
 \label{fig:sp4}
\end{figure}

Performing unquenched simulations with Wilson fermions, extensive
studies of the spectrum (chiral behaviour, discretisation error) have
been presented.  The low-energy description that explicitly
includes the vector and axial-vector states proposed in\cite{Bennett:2017kga}, and based on
the idea of hidden local symmetry\cite{Bando:1984ej}, is used to
determine 10  low energy constants  using the continuum-extrapolated
results of the decay constants and masses in the pseudo-scalar, vector
and axial-vector channels. The effective model describes well
the continuum extrapolated lattice data of the masses and decay
constant squared as a function of the pseudo-scalar mass squared as shown by \fig{fig:sp4}.
 The coupling $g_{\rm{VPP}}$
appears in the effective Lagrangian, and  the value turns
out to be compatible with the KSRF estimate and close to the value
obtained in QCD.  The author argue that it provide empirical support
for the KSRF relation.

\subsection{Toward partial compositeness on the lattice}

As mentioned earlier, an important issue with models of composite
Higgs is to provide mass to the SM  fermions  and in particular to the top
quark.  A mechanism used by phenomenologists in the
context of Composite Higgs models is
referred as partial compositeness. The mechanism provides mass to the
top quark by introducing a coupling between the top
quark and a top-quark partner  at the scale $\mcL_{\rm SCT}$. This effective operator
at the composite scale is generated by an unspecified new sector. A practical
requirement for the PNGB UV completion is therefore to have a spin
$1/2$ bound state that can be 
charged under $SU(3)_c$. UV completion that features  such top partners
have been extensively studied\cite{Ferretti:2013kya,Ferretti:2014qta}.
The authors consider  theories 
involving fermions in mixed representations of the gauge group, and one
of the minimal theory that exhibits top partner candidates is based on
an $SU(4)$ gauge theory with five Weyl fermions in the antisymmetric
representation and three Dirac fermions in the fundamental
representation.

One collaboration undertook instead the extensive
numerical investigation of $SU(4)$ gauge theory with two Dirac fermions in the
antisymmetric representation (denoted $Q$) and two Dirac in the
fundamental (denoted $q$).  Such a gauge theory is not a UV completion
of PNGBs Composite Higgs model, but features spin $1/2$ states ($Qqq$), called
chimera baryons, that 
would correspond to the top partner in the original theory proposed
by \textit{Feretti et al}. In terms of number of Weyl fermions, the simulated
theory is close to Ferreti's UV completion and might serve as a
laboratory to test the mechanism. Any phenomenological conclusion
obtained must therefore acknowledge the strong assumption that the predictions do not change by adding one Weyl fermion
in the antisymmetric representation, and one Dirac in the fundamental.
Several results have been obtained: the masses and decay constants for
the pseudo-scalar and vector mesons\cite{Ayyar:2017qdf},  the baryons
masses constituted of four $q$ operators  and six $Q$ operators, and of
the more interesting $Qqq$ ``chimera'' baryons\cite{Ayyar:2018zuk}.
The necessary chiral extrapolations have been performed using the chiral perturbation
theory of a two-representation theory derived in
\cite{DeGrand:2016pgq}. More recently, another collaboration has
started to investigate the same gauge theory. While they
perform an extensive investigation of the spectrum, the focus is more
on the algorithmic side of the simulations of gauge theories involving
matter content in multiple representations\cite{Cossu:2019hse}.

In summary, while the phenomenological lessons should be taken with
caution, the following results are certainly a first interesting step forward in our
understanding of simulations with mixed representations and of partial compositeness.

\begin{figure}[h!t]
\begin{center}
  \includegraphics[scale=0.5]{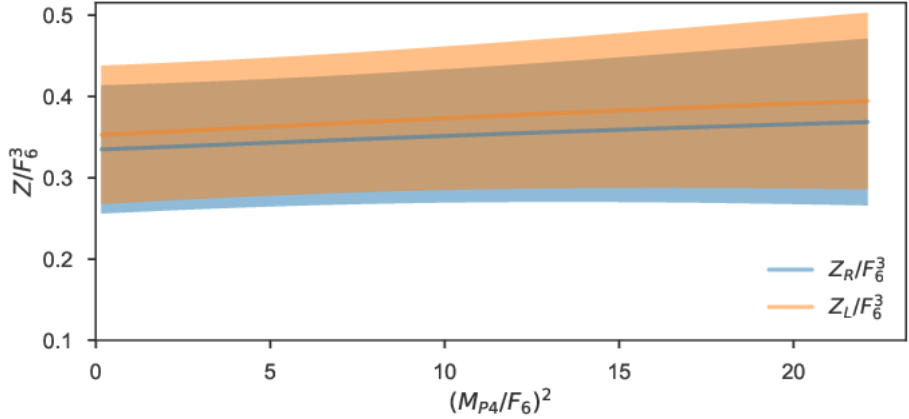}
\caption{\emph{ Chimera baryon matrix elements as a function of  of the quartet mass.}}
\label{fig:partial_compositness}
\end{center}
\end{figure}

The study of the matrix element
relevant in determining the strong sector contribution to the mass of
the top quark has been presented in\cite{Svetitsky:2019hij} and summarises
the recent results obtained in \cite{Ayyar:2018glg}. 
At the scale $\mcL_{\rm SCT}$, the linear mixing
between the chimera baryon and the
massless top quark is controlled by two coupling constant $G_{L,R}$
generated by the extended sector that can be thought as ``Fermi''
constants. At  at $\LEW$ scale, the resulting top Yukawa coupling can be
written :
\begin{align}
y_t \approx G_L G_R \frac{Z_L Z_R}{M_B F_{P6}}
\end{align}
where $M_B F_{P6}$  is a combination of the mass of a chimera baryon and of
pseudo-scalar decay constant of the $Q$ fields,  already computed in
previous studies, and $Z_{L,R}$ are matrix elements of a top partner operator between the
vacuum and the top partner state. The results of the renormalised
matrix element $Z_{L,R} $ is shown in \fig{fig:partial_compositness}.
The renormalisation is performed perturbatively. The authors argue that imposing $y_t \sim 1$, implies that $\LUV$
cannot be much larger than $F_6$ as the ``model'' would require it. We refer to
\cite{Svetitsky:2019hij} and to \cite{Ayyar:2018glg} for more details.

\subsection{Toward lattice constraints of the Higgs potential}
\begin{figure}[h!t]
\begin{center}
\includegraphics[scale=0.5]{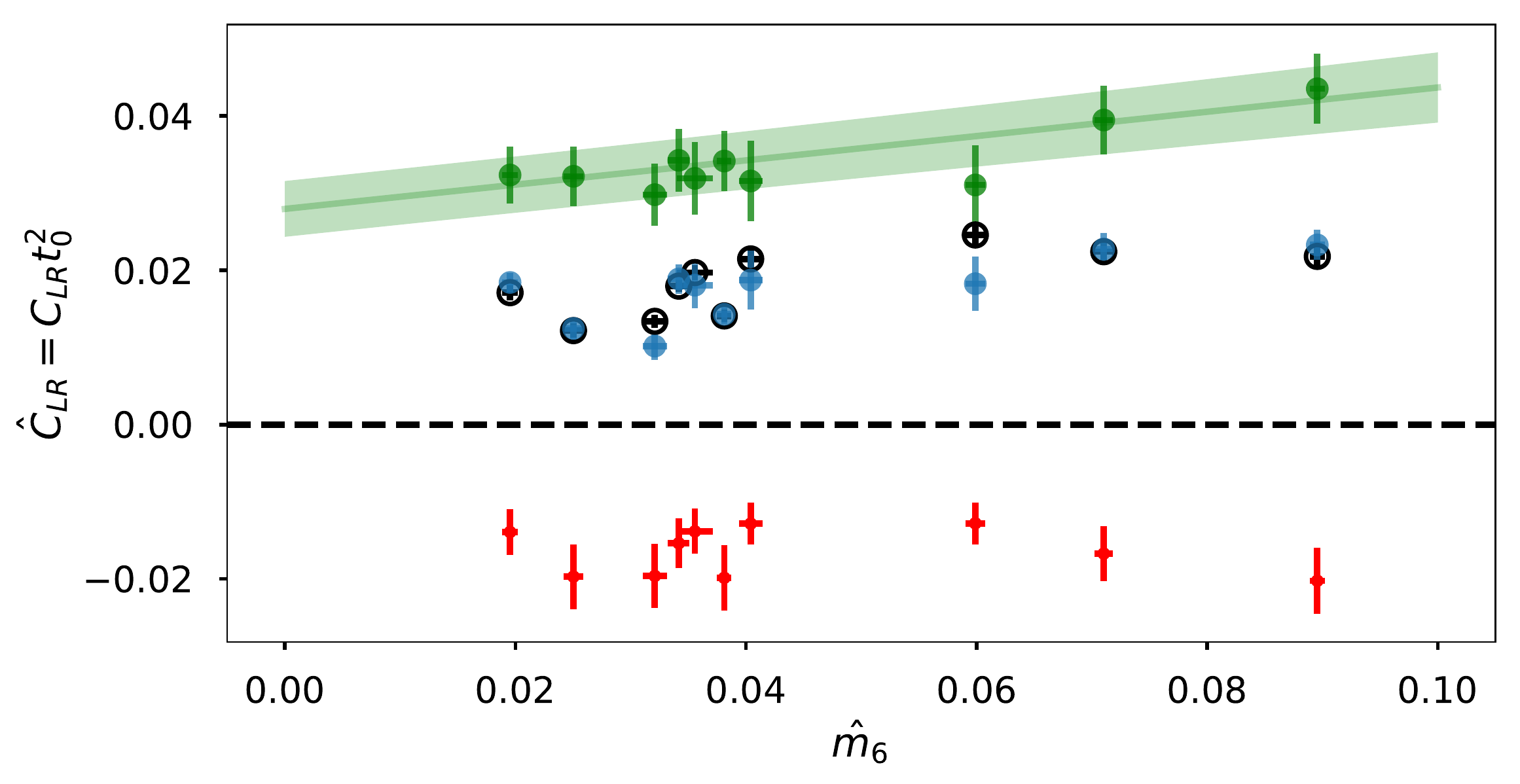}
\caption{\emph{ Fit of the $C_{LR}$ data, the continuum prediction is shown by a green band}}
\label{fig:higgs_pot}
\end{center}
\end{figure}

The Higgs potential is not guaranteed to trigger EWSB: the potential
is induced by loops of gauge bosons and top quarks. The top
contribution requires to estimate a 4-point function, as shown in
\cite{Golterman:2015zwa} and in \cite{DelDebbio:2017ini}. Vector boson
contribution is controlled by a low energy constant, similar to the one that controls
the pion mass spitting in QCD denoted $C_:{LR}$. The low energy
constant has been computed for the first time in \cite{Ayyar:2019exp}
by using a current-current correlator. The authors take the continuum
and chiral limit and find that the value of the low energy constant is
similar to its QCD counterpart in unit of the pseudo-scalar decay
constant. The results  do not allow to
draw any phenomenological conclusion because the theory  is not an UV
completion of a PNGBs model but shows that such calculation is
possible and pave the way to  calculations in more realistic
models. Taking the continuum and chiral limits, the authors of
\cite{Ayyar:2019exp} find that the $C_:{LR}$ in units of the
pseudo-scalar decay constant is roughly of the same size as its QCD
counterpart. The results are summarised as a function of the mass
of the fermions in the antisymmetric representation in \fig{fig:higgs_pot}.

\section{The theory landscape}

We summarise in this section the recent developpement that are less
directly related to the phenomenology but which are crucial to clarify
our understanding of non perturbative gauge dynamics.

\subsection{Conformal Window and near-conformal $\beta$ functions}

A long standing task for the lattice community is to show the
existence of a conformal window, and in particular the critical number
of flavour marking the onset of the conformal window.  The task turns out to be very
challenging. 

Concerning the gauge group $SU(2)$ with fundamental fermions,
extensive investigations suggest no IRFP for $N_f=2,3,4$, while a IRFP
is developed for $N_f=6,7,8$\cite{Amato:2018nvj,Leino:2018yfd}. For adjoint fermions, evidence
for an IRFP have beem found for $N_F=1,3/2,2$ \cite{Athenodorou:2015fda,Rantaharju:2015cne,Rantaharju:2015yva,DelDebbio:2015byq,Bergner:2017gzw} and see
\cite{Bergner:2019kub,Bi:2019gle} for recent developments.

For the $SU(3)$ gauge group, results are controversial. The latest
results for $12$ fundamental fermions, using domain-wall fermions,
support the existence of an IRFP\cite{Hasenfratz:2019puu}. while staggered calculations obtained for
$12$ flavours find no such evidence \cite{Fodor:2019ypi}.  More details on the
latest results can be found in these proceedings. For the antisymmetric
representation, a similar contradiction between lattice results
obtained using the staggered discretisation\cite{Fodor:2019ypi} confirms previous published
results\cite{Fodor:2015zna} and results obtained using Wilson
fermions\cite{Hasenfratz:2015ssa} which find an IRFP. Note that a
an independent spectrum study performed using Wilson fermions is 
compatible with the conformal scenario\cite{Hansen:2017ejh}.

\subsection{Flavour dependence of the ratio $m_V/\fps$}
\begin{figure}[h!t]
\begin{center}
\includegraphics[scale=2.0]{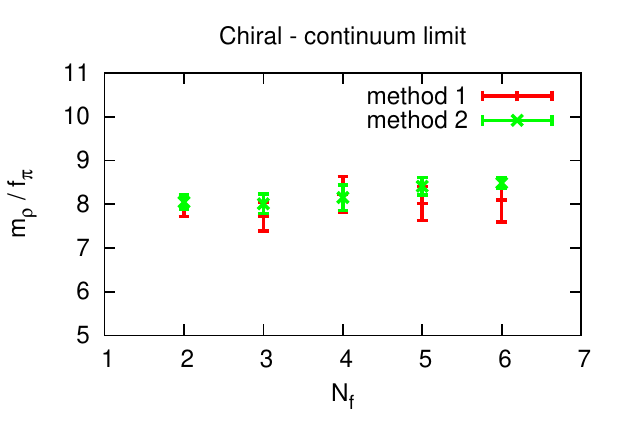}
\caption{\emph{The ratio $m_\rho/f_\pi$ in the chiral-continuum limit
    for each $N_f$ \cite{Nogradi:2019iek}.}}
\label{fig:mrho_over_fpi}
\end{center}
\end{figure}

If a new strongly interacting sector gives rise to a composite Higgs,
the most obvious experimental signature would be the observation of
new composite particles associated to quantum numbers allowed by the
underlying dynamics.  It is expected that, for a large class of
strongly interacting theories, the lightest of such particles would be
the vector meson resonance. It is therefore relevant to determine the
mass of such a state in as many theories as possible. An interesting
work tackled the issue by studying flavour dependence of the ratio
$m_\rho/f_\pi$ below the onset of the conformal windows, i.e in
QCD-like theories. The calculation is performed for the $SU(3)$ gauge
group with $N_f=2,3,4,5,6$ fundamental fermions. 
The mass of the vector meson is obtained from two-point functions in a
regime where the $\rho$ is stable. Continuum extrapolations are
performed for each number of flavours. A careful analysis of the finite volume effects have been performed
and is crucial to obtain the results shown in \fig{fig:mrho_over_fpi}.
The result reads  $m_V/\fps =7.95(15) $ with no significant
$N_f$-dependence. The result relies on extrapolations below the two
pion threshold and on the assumption that ignoring the resonant nature of the
particle does not affect significantly the prediction. 
 The authors also provide a compilation of that ratio
for a number of theories using various gauge groups and fermions in
different representations. Once the leading gauge-group  dependence is
factored out, only a mild gauge group dependence is observed, except
for $SU(2)$ gauge theory.  Assuming that
the KSRF relations holds, such ratio can be used to estimate the value
of the coupling constant $g_{\rho\pi\pi}$. The value obtained suggests
that $g_{\rho\pi\pi}$  is constant when the number of flavour is
changed below the conformal window. Under the assumptions discussed,
the study suggests that the vector resonance coupling is not a
quantity sensitive to the underlying dynamics. The interested
reader is referred to\cite{Nogradi:2019auv} and to a more recent paper \cite{Nogradi:2019iek}.

\subsection{Dynamical generation of particle masses: an alternative to
the Higgs mechanism}

One collaboration is investigating a novel alternative non-perturbative
mechanism for elementary particle mass generation
\cite{Frezzotti:2014wja,Frezzotti:2018zsy}.  The framework is based on
a conjectured non-perturnative  obstruction to the recovery of 
broken fermionic chiral symmetries which give rise to a dynamically
generated fermion mass term. For the first time, the authors provide
numerical evidence of the conjectured phenomenon by using lattice
simulations\cite{Capitani:2019syo}. The authors simulate  within the
simplest 4-dimensional model  in which the phenomona is supposed to
occur: an $SU(3)$ gauge theory with two fundamental
fermions augmented by a colourless scalar doublet, Yukawa terms and a Wilson-like
term.  Tuning the bare parameters to restore the fermionic chiral
symmetry in the effective Lagrangian and performing a continuum extrapolation, they show  that the pseudoscalar
meson mass is significantly different from zero in the phase where the
the exact symmetry acting on fermions and scalars is spontaneously
broken.  
From a phenomenological standpoint, the authors argue the EW
interactions can be included without
introducing tree-level flavour changing neutral currents. The
non-pertubative mechanism would then generate weak bosons mass terms and
provide an alternative to the Higgs mechanism. In such an approach, the
observed Higgs boson would be a bound state of the new interaction appearing  in the vector bosons
scattering channel.  

\section{Conclusions}

The fascinating possibility that composite solutions to the puzzles of
Beyond the Standard Model physics could benefit from first principle
calculations is driving  the lattice
community to study non pertubative phenomena in various gauge
theories. 

In the case of near-conformal dynamics, the presence of light scalars
is a challenge both from the numerical and theoretical points of view.
The studies of PNGBs models is now going beyond spectroscopy and new
calculations provide insight in quantities relevant for the
experiments.

By exploring the dynamics of such theories, the lattice simulations
contribute to deepen our understanding of non-perturbative
effects in Quantum Field Theory.  There is an impressive amount of results on the behavior of quantum field theory,
when the gauge group or the matter content is varied. The results provide quantitative insights which  guide the
model building community to design more realistic models. The lattice
simulations can also test mechanisms relying on non perturbative phenomena in quantum field theory.

Interestingly, the various aspects of the lattice simulations as a
tool to address Physics Beyond the Standard Model raise a number of
challenges. Often, the investigations question the algorithms developped in the context
of Quantum Chromodynamics and their applicability to other theories.
In other cases, the problems require new methods to be developped.
The interpretation of the lattice results often require a thorough and
critical examination in theories that cannot be compared to actual
experimental measures.

\acknowledgments
I would like to thank the organizers for  inviting me and for their
kind hospitality. I would like to thank V. Afferrante, G. Fleming, R. Frezzotti, A. Hasenfratz, K.
Holland, T. Janowski, W. Jay,  J. Kuti, M. Golterman, J.-W. Lee, D. Lin, D. Nogradi, B.
Svetitsky, O. Witzel,  C. H. Wong for useful discussions and for providing
material prior to the conference.
VD received support from the Science and Technology Faculty Council (STFC)
grant  and from the DiRAC data intensive system at the University of
Cambridge and Leicester, operated  on behalf of the U.K. STFC DiRAC
HPC Facility, funded by the Department of Business, Innovation and
Skills national e-infrastructure and STFC capital grants and STFC
Dirac operations grants.

\bibliography{review}
\bibliographystyle{h-physrev}

\end{document}